Explaining the ghosts: Feminist intersectional XAI and cartography as methods to account for invisible labour

Goda Klumbytė

Faculty of Electrical Engineering and Computer Science, University of Kassel, Germany, goda.klumbyte@uni-kassel.de

Hannah Piehl

Faculty of Electrical Engineering and Computer Science, University of Kassel, Germany, hannah.piehl@stud.uni-frankfurt.de

Claude Draude

Faculty of Electrical Engineering and Computer Science, University of Kassel, Germany, claude.draude@uni-kassel.de

Contemporary automation through AI entails a substantial amount of behind-the-scenes human labour, which is often both invisibilised and underpaid. Since invisible labour, including labelling and maintenance work, is an integral part of contemporary AI systems, it remains important to sensitise users to its role. We suggest that this could be done through explainable AI (XAI) design, particularly feminist intersectional XAI. We propose the method of cartography, which stems from feminist intersectional research, to draw out a systemic perspective of AI and include dimensions of AI that pertain to invisible labour.

## 1 INTRODUCTION: EXPLAINABLE AI (XAI) AND INVISIBLE LABOUR

The rise of artificial intelligence has brought forward the need for explaining algorithmic decision-making which is reflected in increasing academic interest in explainable AI (XAI) [1, 2]. While scholars do not agree on a definition of explainability, they all acknowledge gaps in research. What constitutes a (good) explanation still has to be agreed on [1, 3] and the use of terminology in the field of XAI shows a lack of clear distinctions between concepts. For example, *explainability* and *interpretability* are intertwined and often interchangeably used, although scholars mostly agree on interpretability focusing more on the human's ability to make sense of a model and contributing means of information for this to happen, while explainability is seen as a model-centric concept, providing a comprehensible explanation [1, 4].

XAI deals with making the functions of algorithmic models easily understandable to justify their output performance [1, 5]. XAI contains the question "explainable to whom?", suggesting that, ideally, all stakeholders need to be mapped and accounted for [5]. Stakeholder communities include developers, theorists, ethicists, and users [4], all having differing cognitive factors, experience, information needs, as well as various goals, such as trustworthiness, causality, transferability, informativeness, confidence, fairness, accessibility, interactivity or privacy awareness [1, 6]. It is therefore almost impossible to create a model that caters to the requirements of the entire XAI audience.

Invisible labour is an umbrella term that may relate to background work (administrative tasks), routine work (that requires problem-solving skills and advanced knowledge), work by (socially) invisible people (domestic work) or informal work (communicative, social, emotional work) [7]. Feminist scholars have used the concept of invisible work to draw attention to the (often intersecting) gendered, classed, and racialised inequalities and divisions of what is seen as labour (literally and figuratively) and how it is valued [8, 9]. Various scholars have found that when it comes to mapping activities, tasks and affordances of a workplace, only visible and obvious labour is noted [8, 10, 11]. Since ghostly labour, including labelling and maintenance work, is an integral part of contemporary AI systems, it remains crucial to sensitise users to its role and focus on highlighting invisible, undervalued, and underpaid forms of labour.

In academic and professional contexts, *explainability* is used as a technical term, implying that providing explainable algorithmic systems be single-handedly dealt with by technical experts. However, only explaining the workings of a model turns out to not be enough when the system and its effects are not taken into consideration. Instead, the inclusion of a diverse group of stakeholders and other disciplinary knowledge is needed to design XAI systems in order to prevent reproduction of algorithmic bias [12]. This broad view of explainability, we suggest, is the conceptual basis for relying on XAI as a domain that can help generate methods and approaches in HCI and AI that help show AI not as an idealised technical miracle [13] but as a complex technical assemblage and infrastructure that entails human labour and complex power dynamics. To do that fully, however, we suggest XAI needs to additionally draw on feminist epistemological positions to consider the context and situatedness of knowledge making in the XAI process.

## 2 FEMINIST EPISTEMOLOGICAL POSITIONS

Intersectionality illustrates how social categories a person is attributed to or identifies with can intersect and amplify experiences of discrimination or privilege. In addition to race, class and gender, many other social categories – for example religion, sexual orientation, location, dis/ability – are impacting social, cultural and economic resources and notions of power. Using the metaphor of a crossroad, intersectionality serves as an analytical tool to highlight diverse, contextual experiences of discrimination and to show how social categories can interact with and influence each other [14].

Feminist epistemologies argue that knowledge is situated, meaning that there is no such thing as universal, objective, or neutral knowledge [15]. Rather, not unlike social categories, (scientific) knowledge is entangled in social and cultural contexts, establishing knowledge practices that are partial, subjective, and therefore situated. The concept of situated knowledges therefore suggests a more multiple understanding of knowledge through "joining of partial views and halting voices into a collective subject position that promises a vision of the means of ongoing finite embodiment, of living within limits and contradictions – of views from somewhere" [15, p.590]. Standpoint theory describes how a person's standpoint is influencing (scientific) practices of knowledge making and understanding [16, 17, 18, 19, 20, 21]. Understanding knowledge as situated means considering power relations and records of structural, epistemic and systemic violence. By centring marginalised perspectives and drawing attention to minor histories and alternative knowledge practices, often invisible modes of oppression can be made visible and cared for. Standpoint theory not only calls for recognising knowledge in its specific social, cultural, and historical localisations, but advocates for this partiality to be used to counteract the reproduction of bias and discrimination.

## 3 FEMINIST INTERSECTIONAL XAI

Feminist epistemological perspectives expand the framework of XAI by challenging and re-orienting several aspects. First, it requires that explainability would always be understood and designed in specific historical, political, socio-cultural context. Furthermore, because feminist perspective draws attention to intersecting structural inequalities, this context is not limited to immediate socio-technical application setting but includes a broader framework within which the AI system in question is to function. Where exactly the boundaries of the system are drawn will depend on each specific case, however, feminist perspective would necessitate a more structural, systemic and situated understanding of the system [22]. This, contrary to more narrow technical understandings of XAI, would facilitate an integrated approach to XAI [12] and provide a basis to include accounting for invisible forms of labour – data labelling, systems maintenance, infrastructural support – to be included in the scope of systemic operations to be explained.

Second, feminist intersectionality requires paying close attention to power dynamics and centring marginalised perspectives in knowledge making and design practices. Power dynamics in this light concern specifically asking questions: who benefits by the AI system and who is exploited or deprived by it? Who is explaining the system to whom and for what purpose? Furthermore, these questions of power are asked throughout the process of design, and there is a normative imperative here to strive towards more equity and justice and to prioritise perspectives of those who are in positions of less power. Since invisibilised forms of labour, such as Mechanical Turk work, are often underpaid and structured by geopolitical inequalities [23], feminist XAI opens the prospect for such labour and the perspective of workers to be not only addressed but also prioritised as a position for explanation generation.

Third, feminist epistemological stance urges to integrate the question of accountability into knowledge making and design practices. Accountability here is not limited to a narrow perspective of who is legally responsible and who takes the blame when something goes wrong, but as an active effort to foster the capacity to respond: response-ability [15, 24, 25, 26]. Such response-ability necessitates structuring XAI solutions in a way that fosters stakeholders' capacity to critically engage with and respond to the AI system in question. Since the system in feminist perspective is already defined more broadly than a particular functioning machine learning model, invisibilised workers can also be considered as a significant stakeholder group.

To sum up, feminist intersectional XAI, by orienting the field towards contextualisation, attentiveness to power relations and centring of subjugated perspectives, can open a way how to account for ghost work and invisibilised forms of labour and generate explanations that encompass not only explaining specific decisions that AI in question makes, but the functioning of the system and its entanglements with contexts it operates in. We argue that methodologically it can also be helpful to look further into feminist intersectional research for examples of addressing this broader ecosystemic level.

## 4 CARTOGRAPHY AS METHOD

We propose cartography as a useful way to draw out a systemic perspective of AI and include dimensions of AI that pertain to invisibilised labour. We specifically speak here of cartography that is used in feminist cultural studies, feminist philosophy [27, 28, 29, 30, 31] and technoscience [32, 33], where it means a map or tracing of a specific phenomenon with a set of navigational concepts (e.g. gender). The core aspect is its situatedness, i.e., its production from a specific disciplinary, political, cultural positioning. Foregrounding embodied intersectional perspectives, which extends not only to human bodies but also material bodies of technologies and infrastructures that sustain them, it centres subjugated perspectives [27]. Cartographies thus can be intersectional tools to map out AI systems in ways that include invisibilised forms of labor (data collection, labelling, maintenance, etc.) as *integral parts of AI systems* and in this way raise awareness (among the XAI designers as well as end users) of such labor and carework and its power dynamics as constitutive of the functioning of AI systems. For this reason multi-faceted cartographies can be seen as a suitable method for feminist XAI.

Cartographies can be textual, visual, or both. One example of cartography of an AI system addressing and centring invisible labour, is the project "Anatomy of AI" by Crawford and Joeler [34], which provides an overview of human, natural and technical resources, knowledge and operations required to power a smart speaker. In our previous work [35] we used diffractive mapping of a machine learning system to capture and understand the kinds of effects this system might have, entailing an analysis of power relations. In our mapping exercise a team of HCI practitioners focused on understanding these effects through relations of construction, disruption and interference (an example is presented in Figure 1). The participants were asked to indicate these relations in the systems they were analysing by looking closely at their infrastructural level and interaction with social environments. They were specifically encouraged to investigate societal, technical or discipline related elements, discursive or value elements, and the operational logic of the system.

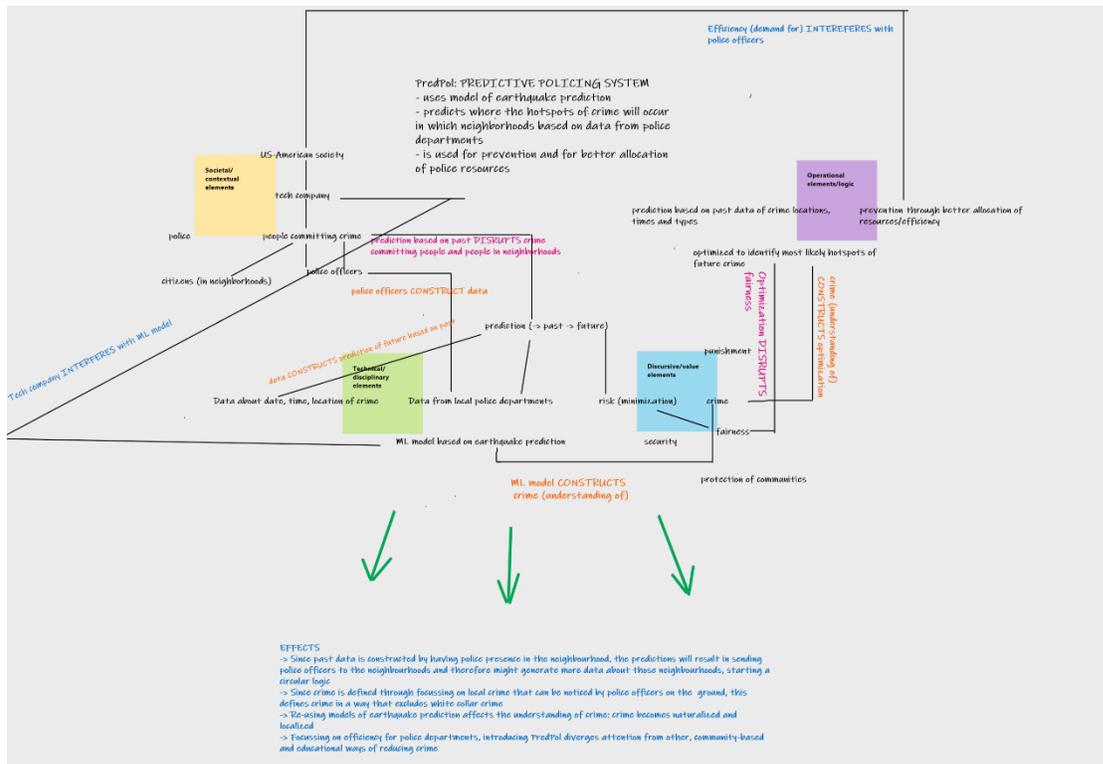

Figure 1: diffractive mapping of predictive policing system PredPol.

While cartography is by no means the only methodological tool to address AI infrastructures and their invisible forms of labour, we suggest it can be a good starting point, particularly for XAI designers, to begin thinking about questions of system interactions with the sociotechnical context, power, and labour.

## 5 DISCUSSION

In this position paper we suggested that feminist intersectional XAI and the method of cartography can help account for invisibilised forms of labour that are powering AI systems. This, hopefully, would also lead to more adequate understanding of AI systems and enable a more productive critical engagement. Further research needs to be carried out on how this conceptual perspective can be operationalised in practice: to what extent could it be operationalised? Should it rather remain as a set of guidelines used for problem definition and sensitisation of designers? Who would be able to operationalise it and what kind of resources and skills would be needed to implement such a more systemic, broader perspective of XAI? For that, we suggest it is important to experiment with interdisciplinary methodologies from social sciences and humanities, and to ask how our collective response-ability as HCI researchers as well as users of AI systems can be fostered towards a more critical engagement.

## ACKNOWLEDGMENTS

This research is supported by Volkswagen Foundation grant "Artificial Intelligence and the Society of the Future" as part of the collaborative project "AI Forensics: Accountability through Interpretability in Visual AI Systems".